\documentclass[superscriptaddress,aps,prl,letter,reprint,showpacs,showkeys,amsmath,amssymb]{revtex4-1}
\usepackage{graphicx}
\usepackage{dcolumn}
\usepackage{bm}
\usepackage{color}
\usepackage{soul}
\usepackage{hyperref}
\usepackage[mathlines]{lineno}

\begin{document}

\preprint{APS/123-QED}

\title{Individual-Ion Addressing with Microwave Field Gradients}

\author{U.~Warring}
\altaffiliation{Present address: Albert-Ludwigs-Universit\"at Freiburg,  Physikalisches Institut, Hermann-Herder-Str. 3, 79104 Freiburg, Germany}
\email{uwarring@gmail.com}
\affiliation{Time and Frequency Division, National Institute of Standards and Technology; 325 Broadway, Boulder, Colorado 80305, USA}

\author{C.~Ospelkaus}
\affiliation{Time and Frequency Division, National Institute of Standards and Technology; 325 Broadway, Boulder, Colorado 80305, USA}\affiliation{QUEST, Leibniz Universit\"at Hannover, Welfengarten 1, 30167 Hannover \\and PTB, Bundesallee 100, 38116 Braunschweig, Germany}

\author{Y.~Colombe}
\author{R.~J\"ordens}
\author{D.~Leibfried}
\email{dil@boulder.nist.gov}
\author{D.~J.~Wineland}
\affiliation{Time and Frequency Division, National Institute of Standards and Technology; 325 Broadway, Boulder, Colorado 80305, USA}

\date{\today}

\begin{abstract}
Individual-qubit addressing is a prerequisite for many instances of quantum information processing. We demonstrate this capability on trapped-ion qubits with microwave near-fields delivered by electrode structures integrated into a microfabricated surface-electrode trap. We describe four approaches that may be used in quantum information experiments with hyperfine levels as qubits. We implement individual control on two $^{25}$Mg$^{+}$ ions separated by $4.3~\mu$m and find spin-flip crosstalk errors on the order of $10^{-3}$.

\end{abstract}


\pacs{37.10.Rs, 37.10.Ty, 03.67.Lx, 32.60.+i}
                              
\maketitle
%

%
Quantum information research is pursued in many physical systems~\cite{ladd_quantum_2010}. Among them, trapped ions are promising for the implementation of qubits and the required logic gates~\cite{cirac_quantum_1995, *blatt_entangled_2008, blatt_quantum_2012}. Previous work has demonstrated elements of an ion-trap array architecture~\citep{rowe_transport_2002, *hensinger_t-junction_2006, *schulz_sideband_2008, *hanneke_realization_2009, amini_toward_2010, *moehring_design_2011, *doret_controlling_2012, blakestad_near-ground-state_2011, walther_controlling_2012, *bowler_coherent_2012} and, by extension, these techniques may be sufficient to perform large-scale quantum computation~\cite{wineland_experimental_1998, *kielpinski_architecture_2002}. Although in most trapped-ion quantum information experiments quantum control is accomplished with lasers~\cite{blatt_entangled_2008, blatt_quantum_2012}, techniques based on microwave fields are also investigated~\cite{mintert_ion-trap_2001, chiaverini_laserless_2008, ospelkaus_trapped-ion_2008, johanning_individual_2009, wang_individual_2009, khromova_designer_2012}. Recently, ion traps incorporating oscillating currents in microfabricated electrode structures have been used for global single-qubit operations~\cite{brown_single-qubit-gate_2011, ospelkaus_microwave_2011, *warring_microwave_2012, allcock_microfabricated_2012} and entangling two-qubit gates~\cite{ospelkaus_microwave_2011, *warring_microwave_2012}. For implementation of universal quantum information processing, this technique requires a novel way to address individual ions from a group and avoid crosstalk. Such addressing methods have been demonstrated with focused laser beams~\cite{naegerl_laser_1999}, differential laser phases~\cite{rowe_experimental_2001}, and static magnetic-field gradients~\cite{schrader_neutral_2004, johanning_individual_2009, *wang_individual_2009}. In this Letter, we describe four methods that use magnetic near fields oscillating at $\simeq 1.7$~GHz to selectively control the spin state of one of two adjacent ions, and characterize crosstalk errors experienced by the unaddressed ion.

\begin{figure}
  \centering
  \includegraphics[width=\columnwidth]{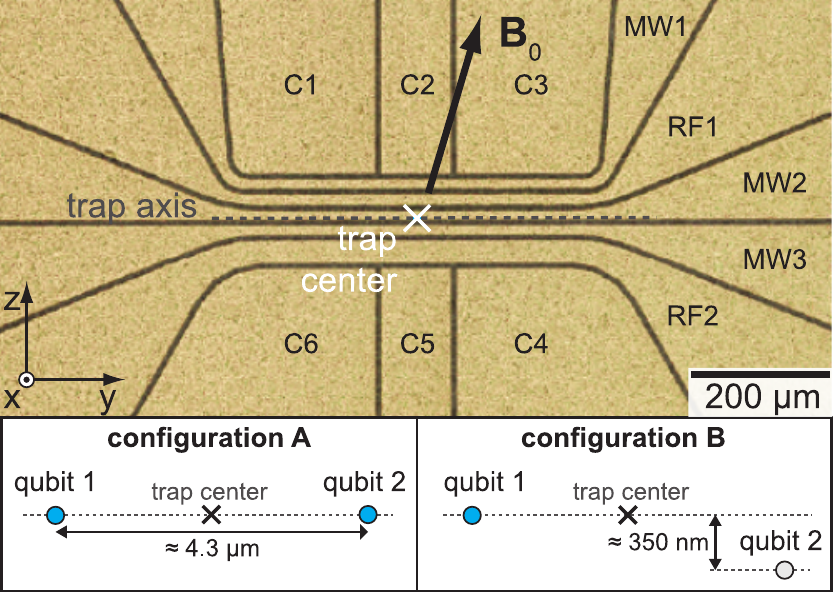}
  \caption{(Color online) Micrograph of the central region of the surface-electrode trap, showing the six control electrodes C1 to C6, the two radio-frequency electrodes RF1 and RF2, and the three microwave electrodes MW1 to MW3. The direction of the external quantization field $|\bold{B_{\text{0}}}| \simeq 21.3$~mT, parallel to the $y$-$z$ plane, is shown. The trap center at $x=y=z=0$ is indicated. (Bottom) Qubit configuration~A is used for global operations including preparation, microwave transfer pulses (with currents in MW2), and detection. Configuration~B, together with currents in all three microwave electrodes, is used for individual-ion addressing (see text).}
\label{fig:method}
\end{figure}

\begin{figure}
  \centering
  \includegraphics[width=\columnwidth]{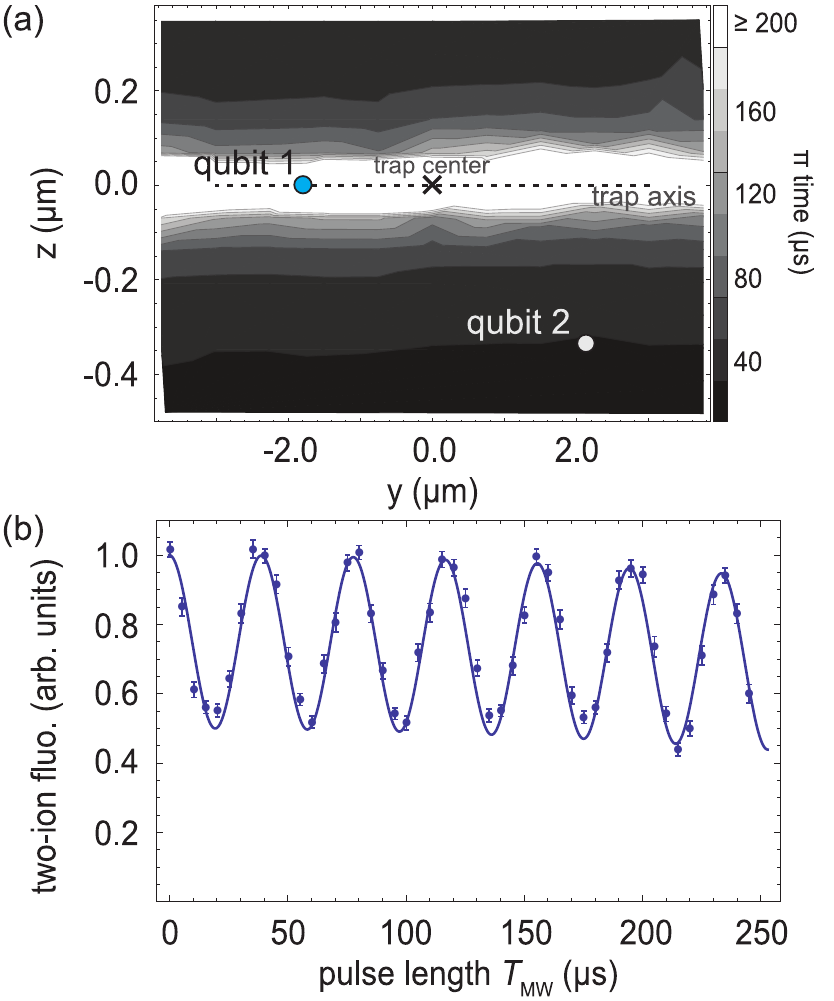}
  \caption{Individual control of two adjacent ion qubits, using method I. (a) Map of $\pi$ times as a function of ion position in the plane parallel to the trap surface. The Rabi rate is probed with a single ion in 120 positions (with relative precision $< 2~\%$), and the data points are interpolated to illustrate the spatial variation. Ion positions in Configuration~B are indicated. (b) Two-ion detection fluorescence trace  (proportional to  $P(\downarrow,1) + P(\downarrow,2)$, where $P(\downarrow,i)$ is the probability of ion $i$ in state $\left| \downarrow \right>$) as a function of the microwave pulse length in Configuration~B.} 
\label{fig:flopping}
\end{figure}

The experiments use two $^{25}$Mg$^+$ ions confined in a surface-electrode Paul trap~\cite{seidelin_microfabricated_2006} at a distance $d \simeq 30~\mu$m above the surface; details of the apparatus are given in~\cite{warring_microwave_2012}. The trap incorporates six electro-static control electrodes, two radio-frequency electrodes driven at $\omega_{\text{RF}} \simeq 2\pi \times 71.6$~MHz, and three microwave electrodes for generating oscillating magnetic near-fields (Fig.~\ref{fig:method}). Typical single-ion motional mode frequencies are $\omega_{\text{axial}} \simeq 2\pi \times 1.4$~MHz in the $y$ (axial) direction and $\omega_{\text{radial}} \simeq 2\pi \times 7.0$~MHz in the $x$-$z$ (radial) plane. For these experimental parameters two Mg$^+$ ions align along the $y$ axis with an inter-ion spacing of $\simeq 4.3~\mu$m.
The quantization axis is defined by a static magnetic field $|\bold{B}_{0}| \simeq 21.3$~mT (produced by external coils) parallel to the trap surface and at an angle of $15^{\circ}$ with respect to the $z$ axis. At this field strength the $\left|F = 3, m_{\text{F}} = 1\right> \equiv \left|\downarrow\right>$ to $\left|F = 2,m_{\text{F}} = 1\right>\equiv \left|\uparrow\right>$ hyperfine-qubit transition~\footnote{Here $F$ is the total angular momentum and $m_{\text{F}}$ is the projection of the angular momentum onto the magnetic field axis} at $\omega_{\text{q}} \simeq 2\pi \times1.687$~GHz is to first order field-independent ($\delta \omega_{\text{q}}/ \delta |\bold{B}_{0}| = 0$). Such transitions are favorable because of their long coherence times~\cite{langer_long-lived_2005}. To initialize the experiment, the ions are Doppler cooled and optically pumped to the $\left| 3, 3\right>$ ground state by two superimposed $\sigma^{+}$-laser beams parallel to $\bold{B}_{0}$ tuned nearly resonant with the $^2$S$_{1/2} \left|3,3\right> \rightarrow ^2$P$_{3/2}\left|4,4\right>$ cycling transition~\cite{warring_microwave_2012}. Two sequential global hyperfine-state transfer $\pi$ pulses implemented with microwave currents in electrode MW2 populate the $\left| \downarrow \right>$ state of the qubits. For detection, to discriminate $\left| \downarrow \right>$ from $\left| \uparrow \right>$, we first reverse this process, transferring $\left| \downarrow \right>$ to $\left|3,3\right>$ and apply similar pulses to transfer $\left| \uparrow \right>$ to $\left|2,-1\right>$. We then excite the ions on the cycling transition to indicate their internal state (cp. Fig.\ref{fig:flopping}b).

Individual qubit control is accomplished by selective positioning of the ions in a spatially varying microwave magnetic field $\bold{B}_{\text{MW}}(x, y, z)$ that oscillates at frequency $\omega_{\text{MW}}$. Near the center of the trap $\bold{B}_{\text{MW}}$ can be approximated for $\sqrt{x^2+ z^2} \lesssim 3~\mu$m by a $y$-independent $x$-$z$ quadrupole field. Currents in all three microwave electrodes are adjusted to generate microwave fields with $\left|\bold{B}_{\text{MW}} \right| \simeq 0$ on the trap axis and gradients between $7$~T/m and $35$~T/m in the radial plane~\cite{ospelkaus_microwave_2011, *warring_microwave_2012}. We apply control potentials to place the ions in configurations A or B as illustrated in Fig.~\ref{fig:method}. Configuration~A, where both ions are on the trap axis, is used for global operations: state preparation and detection, and, with currents in MW2, for common qubit operations. Configuration~B, where ion 2 is shifted $\simeq 350$~nm off axis, together with currents in all microwave electrodes, enables the individual addressing of qubit~2. We adiabatically switch between the two configurations in $\simeq 80~\mu$s.

In method I, qubit 2 is driven on resonance by $\bold{B}_{\text{MW}}$ while the field strength is minimal at the position of qubit~1. The qubit transition is driven by $B_\|$, the component of $\bold{B}_{\text{MW}}$ parallel to $\bold{B}_{0}$. We configure $\left |\bold{B}_{\text{MW}}(0, y, 0) \right| \simeq 0$ at $\omega_{\text{MW}} = \omega_{\text{q}}$ as described in~\cite{ospelkaus_microwave_2011, *warring_microwave_2012} and, with a single ion, we map the qubit $\pi$ time $T_{\pi,\text{q}}(0, y, z) \propto B^{-1}_\|(0, y, z)$ as a function of position (Fig.~\ref{fig:flopping}a).
From a model fit to this map, we find a $\delta B_\|/\delta z = 7.1(5)$~T/m and a residual $B_{\|}(0,y,0) = 0.14(1)~\mu$T. To demonstrate individual addressing, two qubits are initialized in $\left| \downarrow \downarrow \right>$ while being held in configuration~A. The ion positions are then shifted to configuration~B. After applying $\bold{B}_{\text{MW}}$ for duration $T_{\text{MW}}$ the positions are switched back to configuration~A and the qubit states of both ions are detected (Fig.~\ref{fig:flopping}b). The Rabi rates $\Omega_{\text{q1}} =2\pi \times 0.32(2)$~kHz and $\Omega_{\text{q2}} = 2\pi \times 12.84(6)$~kHz are extracted from a model fit to the data. For an applied $\pi$ pulse on qubit~2, the spin-flip probability (which we refer to as crosstalk error) of qubit~1 is $1.5(2)\times 10^{-3}$. The suppression of $\Omega_{\text{q1}}$ is limited by the accuracy of individual phase and amplitude control of the currents fed into the three microwave electrodes~\cite{warring_microwave_2012}.

Method~II is based on the approach presented in~\cite{turchette_deterministic_1998, *leibfried_individual_1999}; the displacement of ion 2 causes excess micromotion, which enables the addressing on the radio-frequency micromotion sideband~\cite{berkeland_minimization_1998}. The corresponding Rabi rate $\Omega_{\text{mm}}$ is proportional to $\bold{r}_{\text{mm}}\cdot \nabla B_\|$, where $|\bold{r}_{\text{mm}}|$ is the micromotion amplitude~\cite{ospelkaus_trapped-ion_2008}. We apply a gradient $\delta B_\|/\delta z \simeq 35$~T/m at $\omega_{\text{MW}} = \omega_{\text{q}} - \omega_{\text{RF}}$ and minimize the field on the trap axis as in method I, to avoid large ac Zeeman shifts. We measure Rabi rates $\Omega_{\text{mm,q1}} = 2\pi \times 0.05(1)$~kHz and $\Omega_{\text{mm,q2}} = 2\pi \times 3.11(2)$~kHz, corresponding to a crosstalk error of $6(3)\times 10^{-4}$. The residual micromotion amplitude $0.42(6)$~nm of ion 1 may be limited by the positioning precision and/or unequal phases of the radio-frequency electrodes~\cite{berkeland_minimization_1998}. This method leads to a differential ac Zeeman shift $\delta \omega_{\text{acz}} \simeq 2\pi \times 430$~Hz, due to oscillating field amplitudes $|\bold{B}_{\text{MW,q1}}| \simeq 7~\mu$T and $|\bold{B}_{\text{MW,q2}}| \simeq 19~\mu$T, which must be compensated.

Method III is based on differential ac Zeeman shifts on the ions, which gives differential $\sigma_{\text{z}}$ control. Together with global operations, this enables full individual control and is analogous to the addressing approach based on differential ac Stark shifts~\cite{blatt_quantum_2012}. Here, $\bold{B}_{\text{MW}}$ is applied at $\omega_{\text{MW}} = \omega_{\text{q}} + \Delta$, where the detuning $\Delta$ induces a spatially varying ac Zeeman shift $\omega_{\text{acz}} = c_\| B^2_\| + c_\perp B^2_\perp$, where $B_\perp$ is the component of $\bold{B}_{\text{MW}}$ perpendicular to $\bold{B}_0$. The coefficients $c_\|$ and $c_\perp$ depend on $\Delta$ and can be calculated from the relevant Clebsch-Gordan coefficients~\cite{warring_microwave_2012}. Any $\sigma_{\text{z}}$ rotation on qubit~1 can be accounted for in subsequent computations, or suppressed by applying a compensating ac Zeeman shift; the crosstalk is limited by the degree to which the $\sigma_{\text{z}}$ phase is determined.

\begin{figure}
  \centering
  \includegraphics[width=\columnwidth]{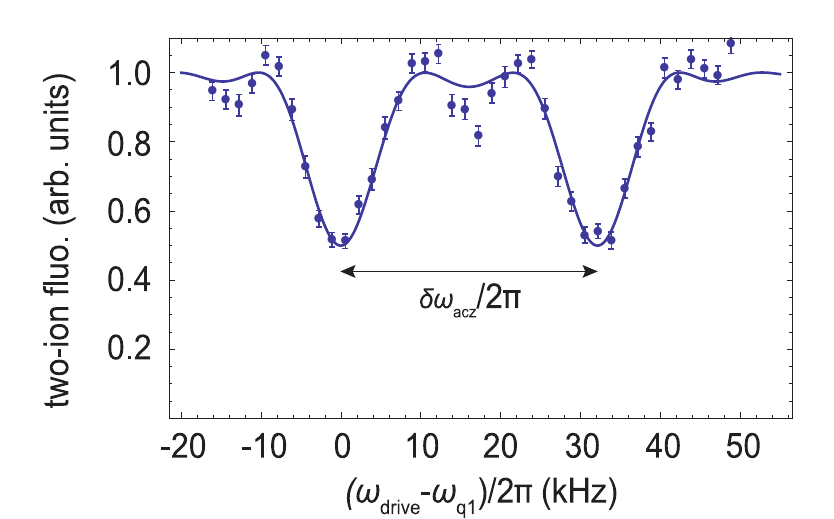}
  \caption{Individual control using method IV. Two-ion detection fluorescence (proportional to $P(\downarrow,1) + P(\downarrow,2))$ as a function of drive frequency $\omega_{\text{drive}}$ applied to MW2. Here, $\omega_{\text{q1}}$ is the resonance frequency of qubit~1. The drive duration is set to apply a $\pi$ pulse on the qubits when on resonance. The qubit resonances are separated by a differential ac Zeeman shift $\delta \omega_{\text{acz}}= 2\pi \times 32.1(3)$~kHz.} 
\label{fig:IIIandIV}
\end{figure}

Method IV extends method III: the spatially varying ac Zeeman shift splits the qubit resonances by $\delta \omega_{\text{acz}}$ and a drive signal on MW2 addresses the qubits. This drive field will lead to approximately the same resonant Rabi rate $\Omega_{\text{q}}$ for both qubits. For the experiment we choose $\Delta \simeq - 2\pi \times 3.0$~MHz. We observe a separation $\delta \omega_{\text{acz}} = 2\pi \times 32.1(3)$~kHz between the qubit transitions (Fig.~\ref{fig:IIIandIV}). For $\Omega_{\text{q}} = 2\pi \times2.08(2)$~kHz the crosstalk, given by the probability of off-resonant transitions, is $1.1(9)\times 10^{-3}$. The differential ac Zeeman shift must be accounted for in subsequent operations. Since $\Omega_{\text{q}} < |\delta \omega_{\text{acz}}|$, this method is slower than method III.

\begin{figure}
  \centering
  \includegraphics[width=\columnwidth]{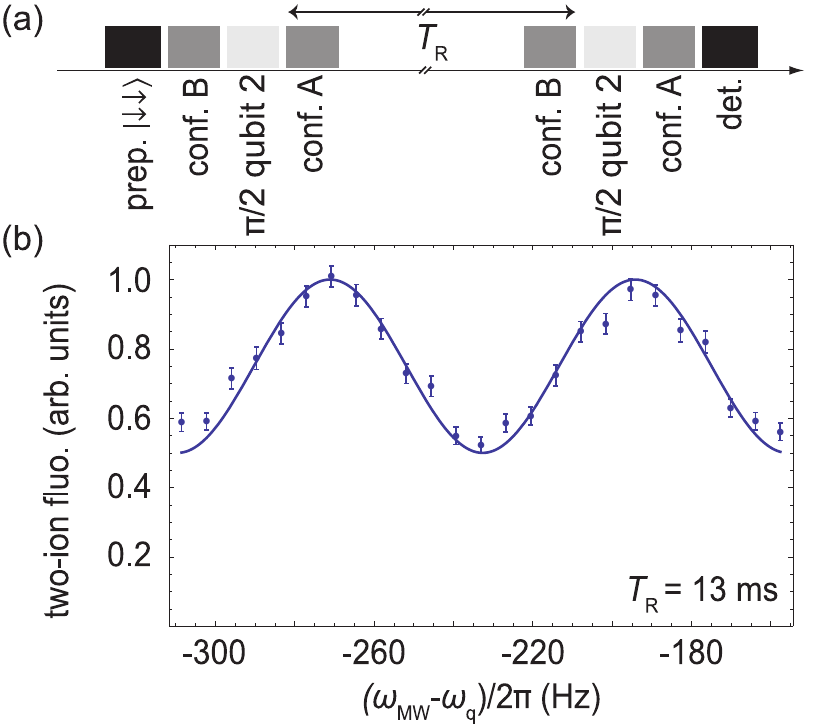}
  \caption{Ramsey experiment on qubit~2 using method I. (a) Pulse sequence: qubits~1 and 2 are initialized into the $\left|\downarrow\downarrow\right>$ state. The ions are then placed in configuration~B, a $\pi / 2$ pulse is applied to qubit~2, and the ions are switched back into configuration~A. After a wait duration $T_{\text{R}}$, the ions are moved to configuration B and a second $\pi/2$ pulse is applied to qubit~2 followed by fluorescence detection of both ions in configuration~A. (b) Two-ion detection fluorescence as a function of the $\pi/2$ pulse frequency for $T_{\text{R}} = 13$~ms. Ramsey fringes of qubit~2 are visible, while qubit~1 remains in $\left|\downarrow\right>$ giving rise to the overall fluorescence offset.} 
\label{fig:ramsey}
\end{figure}

To determine the effect of spatial reconfiguration on qubit coherence, we perform two types of Ramsey experiments and observe the decrease in Ramsey fringe contrast as a function of free-precession time $T_{\text{R}}$. In a reference experiment, we observe a qubit coherence time~\footnote{The coherence time is defined as the Ramsey interval $T_{\text{R}}$ at which the fringe contrast decays by a factor $e^{-1}$.} longer than $200$~ms for a single ion located at the trap center while keeping the control potentials constant. Here, two $\pi/2$ pulses, separated by time $T_{\text{R}}$, are applied with the global microwave drive. In a second experiment, we prepare two qubits in $\left| \downarrow \downarrow \right>$ and perform a $\pi/2$ pulse on qubit~2 using method~I (Fig.~\ref{fig:ramsey}a). 
The ion positions are then switched back to configuration~A, and after $T_{\text{R}}$ a second $\pi/2$ pulse is applied to qubit~2. Subsequently the two-ion fluorescence is detected. Figure~\ref{fig:ramsey}b shows results for $T_{\text{R}} = 13$~ms. We observe no additional loss in contrast due to the repositioning of the ions. However, in both experiments, a precise measurement and comparison is hampered by a significant loss ($\simeq 50~\%$) of overall fluorescence due to motional heating of the ion(s) after $\simeq 50$~ms without laser cooling.

\begin{table}
\caption{Comparison of individual addressing methods. $\Omega_{\text{q1}}$ and $\Omega_{\text{q2}}$ denote the individual qubit Rabi rates and for method III they denote $\sigma_{\text{z}}$-rotation rates. The crosstalk error is the probability of a spin flip on qubit~1 when applying a $\pi$ pulse on qubit~2. The differential ac Zeeman shift is absent in method I. Crosstalk for method III depends on the degree to which the phase shift on qubit 1 can be compensated.}
\begin{center}
\begin{tabular}{ccccc}
\hline
\hline
method & $\Omega_{\text{q1}}/(2\pi)$ & $\Omega_{\text{q2}}/(2\pi)$ & crosstalk & $\delta\omega_{\text{acz}}/(2\pi)$\\
 & (kHz) & (kHz) &  ($\times 10^{-3}$) & (kHz)\\
\hline
I& $0.32(2)$ & $12.84(6) $ & $1.5(2)$ & --\\
II&$0.05(1)$ & $3.11(2) $ & $0.6(3)$ & $\simeq 0.43$\\
III&$\simeq 4.7$&$\simeq 36.8$& -- & $32.1(3)$\\
IV& $2.08(2)$ & $2.08(2)$ & $1.1(9)$ & $32.1(3)$\\
\hline
\hline
\end{tabular}
\end{center}
\label{tab:comp}
\end{table}

In conclusion, we have demonstrated four methods for individual addressing of two qubits by use of microwave near-field gradients. These methods may enable a processor architecture that is based only on oscillating near-fields for coherent ion qubit control. A summary of their overall performance is listed in Table~\ref{tab:comp}. Crosstalk and Rabi rates are currently limited by ion position control as well as relative phase and amplitude control of the signals driving the three microwave electrodes. All schemes can be augmented by pulse shaping and composite pulse schemes~\cite{levitt_composite_1986}.

In future applications, these methods may be implemented in a linear trap array where ions reside in separated potential wells. In this case, the addressed ion can be pushed much farther away ($\geq 1~\mu$m) from the trap axis, decreasing the crosstalk. For faster switching of control potentials diabatic methods can be used~\cite{walther_controlling_2012, *bowler_coherent_2012}. The addressing methods can also be used for individual detection of multiple ions stored in the same potential wells by appropriate detection sequences. For example a detection sequence for two qubits could consist of two consecutive detection pulses separated by a $\pi$ pulse on qubit 2. Here, the $\pi$ pulse and second detection pulse are necessary only when the first detection results in one ``bright'' ($\left| \downarrow \right>$ state) and one ``dark'' ($\left| \uparrow \right>$ state) ion. Table~\ref{tab:det} shows a truth table to illustrate the possible detection outcomes.
\begin{table}
\caption{Truth table for the conditional detection sequence of two ions. Combinations of qubit states and the corresponding number of ions detected as bright are listed. The $\pi$ pulse and detection 2 need to be performed only when detection 1 indicates one bright ion.}
\begin{center}
\begin{tabular}{cccc}
\hline
\hline
qubit 1 &qubit 2 &detection 1 &detection 2 \\
\hline
$\left| \downarrow \right>$ & $\left| \downarrow \right>$ & 2 & -- \\
$\left| \downarrow \right>$ & $\left| \uparrow \right>$ & 1 & 2 \\
$\left| \uparrow \right>$ & $\left| \downarrow \right>$ & 1 & 0 \\
$\left| \uparrow \right>$ & $\left| \uparrow \right>$ & 0 & -- \\
\hline
\hline
\end{tabular}
\end{center}
\label{tab:det}
\end{table}

While preparing this manuscript, we became aware of a related experiment that uses laser fields and differential micromotion to enable single-ion addressing~\cite{navon_single-spin_2012}.
\begin{acknowledgments}
We thank K. R. Brown and J. M. Amini  for their support in designing and building parts of the apparatus, and S. M. Brewer and J. Britton for helpful comments on the manuscript. This work was supported by IARPA, ARO contract \#EAO139840, ONR, DARPA, Sandia National Laboratories and the NIST Quantum Information Program. Contribution of NIST, not subject to U.~S.~copyright.
\end{acknowledgments}


\begin{thebibliography}{38}%
\makeatletter
\providecommand \@ifxundefined [1]{%
 \@ifx{#1\undefined}
}%
\providecommand \@ifnum [1]{%
 \ifnum #1\expandafter \@firstoftwo
 \else \expandafter \@secondoftwo
 \fi
}%
\providecommand \@ifx [1]{%
 \ifx #1\expandafter \@firstoftwo
 \else \expandafter \@secondoftwo
 \fi
}%
\providecommand \natexlab [1]{#1}%
\providecommand \enquote  [1]{``#1''}%
\providecommand \bibnamefont  [1]{#1}%
\providecommand \bibfnamefont [1]{#1}%
\providecommand \citenamefont [1]{#1}%
\providecommand \href@noop [0]{\@secondoftwo}%
\providecommand \href [0]{\begingroup \@sanitize@url \@href}%
\providecommand \@href[1]{\@@startlink{#1}\@@href}%
\providecommand \@@href[1]{\endgroup#1\@@endlink}%
\providecommand \@sanitize@url [0]{\catcode `\\12\catcode `\$12\catcode
  `\&12\catcode `\#12\catcode `\^12\catcode `\_12\catcode `\%12\relax}%
\providecommand \@@startlink[1]{}%
\providecommand \@@endlink[0]{}%
\providecommand \url  [0]{\begingroup\@sanitize@url \@url }%
\providecommand \@url [1]{\endgroup\@href {#1}{\urlprefix }}%
\providecommand \urlprefix  [0]{URL }%
\providecommand \Eprint [0]{\href }%
\providecommand \doibase [0]{http://dx.doi.org/}%
\providecommand \selectlanguage [0]{\@gobble}%
\providecommand \bibinfo  [0]{\@secondoftwo}%
\providecommand \bibfield  [0]{\@secondoftwo}%
\providecommand \translation [1]{[#1]}%
\providecommand \BibitemOpen [0]{}%
\providecommand \bibitemStop [0]{}%
\providecommand \bibitemNoStop [0]{.\EOS\space}%
\providecommand \EOS [0]{\spacefactor3000\relax}%
\providecommand \BibitemShut  [1]{\csname bibitem#1\endcsname}%
\let\auto@bib@innerbib\@empty
\bibitem [{\citenamefont {Ladd}\ \emph {et~al.}(2010)\citenamefont {Ladd},
  \citenamefont {Jelezko}, \citenamefont {Laflamme}, \citenamefont {Nakamura},
  \citenamefont {Monroe},\ and\ \citenamefont {{O'Brien}}}]{ladd_quantum_2010}%
  \BibitemOpen
  \bibfield  {author} {\bibinfo {author} {\bibfnamefont {T.~D.}\ \bibnamefont
  {Ladd}}, \bibinfo {author} {\bibfnamefont {F.}~\bibnamefont {Jelezko}},
  \bibinfo {author} {\bibfnamefont {R.}~\bibnamefont {Laflamme}}, \bibinfo
  {author} {\bibfnamefont {Y.}~\bibnamefont {Nakamura}}, \bibinfo {author}
  {\bibfnamefont {C.}~\bibnamefont {Monroe}}, \ and\ \bibinfo {author}
  {\bibfnamefont {J.~L.}\ \bibnamefont {{O'Brien}}},\ }\href@noop {} {\bibfield
   {journal} {\bibinfo  {journal} {Nature}\ }\textbf {\bibinfo {volume}
  {464}},\ \bibinfo {pages} {45} (\bibinfo {year} {2010})}\BibitemShut
  {NoStop}%
\bibitem [{\citenamefont {Cirac}\ and\ \citenamefont
  {Zoller}(1995)}]{cirac_quantum_1995}%
  \BibitemOpen
  \bibfield  {author} {\bibinfo {author} {\bibfnamefont {J.~I.}\ \bibnamefont
  {Cirac}}\ and\ \bibinfo {author} {\bibfnamefont {P.}~\bibnamefont {Zoller}},\
  }\href@noop {} {\bibfield  {journal} {\bibinfo  {journal} {Phys. Rev. Lett.}\
  }\textbf {\bibinfo {volume} {74}},\ \bibinfo {pages} {4091} (\bibinfo {year}
  {1995})}\BibitemShut {NoStop}%
\bibitem [{\citenamefont {Blatt}\ and\ \citenamefont
  {Wineland}(2008)}]{blatt_entangled_2008}%
  \BibitemOpen
  \bibfield  {author} {\bibinfo {author} {\bibfnamefont {R.}~\bibnamefont
  {Blatt}}\ and\ \bibinfo {author} {\bibfnamefont {D.}~\bibnamefont
  {Wineland}},\ }\href@noop {} {\bibfield  {journal} {\bibinfo  {journal}
  {Nature}\ }\textbf {\bibinfo {volume} {453}},\ \bibinfo {pages} {1008}
  (\bibinfo {year} {2008})}\BibitemShut {NoStop}%
\bibitem [{\citenamefont {Blatt}\ and\ \citenamefont
  {Roos}(2012)}]{blatt_quantum_2012}%
  \BibitemOpen
  \bibfield  {author} {\bibinfo {author} {\bibfnamefont {R.}~\bibnamefont
  {Blatt}}\ and\ \bibinfo {author} {\bibfnamefont {C.~F.}\ \bibnamefont
  {Roos}},\ }\href@noop {} {\bibfield  {journal} {\bibinfo  {journal} {Nat.
  Phys.}\ }\textbf {\bibinfo {volume} {8}},\ \bibinfo {pages} {277} (\bibinfo
  {year} {2012})}\BibitemShut {NoStop}%
\bibitem [{\citenamefont {Rowe}\ \emph {et~al.}(2002)\citenamefont {Rowe},
  \citenamefont {Ben-Kish}, \citenamefont {{DeMarco}}, \citenamefont
  {Leibfried}, \citenamefont {Meyer}, \citenamefont {Beall}, \citenamefont
  {Britton}, \citenamefont {Hughes}, \citenamefont {Itano}, \citenamefont
  {Jelenkovi{\'c}}, \citenamefont {Langer}, \citenamefont {Rosenband},\ and\
  \citenamefont {Wineland}}]{rowe_transport_2002}%
  \BibitemOpen
  \bibfield  {author} {\bibinfo {author} {\bibfnamefont {M.~A.}\ \bibnamefont
  {Rowe}}, \bibinfo {author} {\bibfnamefont {A.}~\bibnamefont {Ben-Kish}},
  \bibinfo {author} {\bibfnamefont {B.}~\bibnamefont {{DeMarco}}}, \bibinfo
  {author} {\bibfnamefont {D.}~\bibnamefont {Leibfried}}, \bibinfo {author}
  {\bibfnamefont {V.}~\bibnamefont {Meyer}}, \bibinfo {author} {\bibfnamefont
  {J.}~\bibnamefont {Beall}}, \bibinfo {author} {\bibfnamefont
  {J.}~\bibnamefont {Britton}}, \bibinfo {author} {\bibfnamefont
  {J.}~\bibnamefont {Hughes}}, \bibinfo {author} {\bibfnamefont {W.~M.}\
  \bibnamefont {Itano}}, \bibinfo {author} {\bibfnamefont {B.}~\bibnamefont
  {Jelenkovi{\'c}}}, \bibinfo {author} {\bibfnamefont {C.}~\bibnamefont
  {Langer}}, \bibinfo {author} {\bibfnamefont {T.}~\bibnamefont {Rosenband}}, \
  and\ \bibinfo {author} {\bibfnamefont {D.~J.}\ \bibnamefont {Wineland}},\
  }\href@noop {} {\bibfield  {journal} {\bibinfo  {journal} {Quantum Inf.
  Comput.}\ }\textbf {\bibinfo {volume} {2}},\ \bibinfo {pages} {257} (\bibinfo
  {year} {2002})}\BibitemShut {NoStop}%
\bibitem [{\citenamefont {Hensinger}\ \emph {et~al.}(2006)\citenamefont
  {Hensinger}, \citenamefont {Olmschenk}, \citenamefont {Stick}, \citenamefont
  {Hucul}, \citenamefont {Yeo}, \citenamefont {Acton}, \citenamefont
  {Deslauriers}, \citenamefont {Monroe},\ and\ \citenamefont
  {Rabchuk}}]{hensinger_t-junction_2006}%
  \BibitemOpen
  \bibfield  {author} {\bibinfo {author} {\bibfnamefont {W.~K.}\ \bibnamefont
  {Hensinger}}, \bibinfo {author} {\bibfnamefont {S.}~\bibnamefont
  {Olmschenk}}, \bibinfo {author} {\bibfnamefont {D.}~\bibnamefont {Stick}},
  \bibinfo {author} {\bibfnamefont {D.}~\bibnamefont {Hucul}}, \bibinfo
  {author} {\bibfnamefont {M.}~\bibnamefont {Yeo}}, \bibinfo {author}
  {\bibfnamefont {M.}~\bibnamefont {Acton}}, \bibinfo {author} {\bibfnamefont
  {L.}~\bibnamefont {Deslauriers}}, \bibinfo {author} {\bibfnamefont
  {C.}~\bibnamefont {Monroe}}, \ and\ \bibinfo {author} {\bibfnamefont
  {J.}~\bibnamefont {Rabchuk}},\ }\href@noop {} {\bibfield  {journal} {\bibinfo
   {journal} {Appl. Phys. Lett.}\ }\textbf {\bibinfo {volume} {88}},\ \bibinfo
  {pages} {034101} (\bibinfo {year} {2006})}\BibitemShut {NoStop}%
\bibitem [{\citenamefont {Schulz}\ \emph {et~al.}(2008)\citenamefont {Schulz},
  \citenamefont {Poschinger}, \citenamefont {Ziesel},\ and\ \citenamefont
  {{Schmidt-Kaler}}}]{schulz_sideband_2008}%
  \BibitemOpen
  \bibfield  {author} {\bibinfo {author} {\bibfnamefont {S.~A.}\ \bibnamefont
  {Schulz}}, \bibinfo {author} {\bibfnamefont {U.}~\bibnamefont {Poschinger}},
  \bibinfo {author} {\bibfnamefont {F.}~\bibnamefont {Ziesel}}, \ and\ \bibinfo
  {author} {\bibfnamefont {F.}~\bibnamefont {{Schmidt-Kaler}}},\ }\href@noop {}
  {\bibfield  {journal} {\bibinfo  {journal} {New J. Phys.}\ }\textbf {\bibinfo
  {volume} {10}},\ \bibinfo {pages} {045007} (\bibinfo {year}
  {2008})}\BibitemShut {NoStop}%
\bibitem [{\citenamefont {Hanneke}\ \emph {et~al.}(2010)\citenamefont
  {Hanneke}, \citenamefont {Home}, \citenamefont {Jost}, \citenamefont {Amini},
  \citenamefont {Leibfried},\ and\ \citenamefont
  {Wineland}}]{hanneke_realization_2009}%
  \BibitemOpen
  \bibfield  {author} {\bibinfo {author} {\bibfnamefont {D.}~\bibnamefont
  {Hanneke}}, \bibinfo {author} {\bibfnamefont {J.~P.}\ \bibnamefont {Home}},
  \bibinfo {author} {\bibfnamefont {J.~D.}\ \bibnamefont {Jost}}, \bibinfo
  {author} {\bibfnamefont {J.~M.}\ \bibnamefont {Amini}}, \bibinfo {author}
  {\bibfnamefont {D.}~\bibnamefont {Leibfried}}, \ and\ \bibinfo {author}
  {\bibfnamefont {D.~J.}\ \bibnamefont {Wineland}},\ }\href@noop {} {\bibfield
  {journal} {\bibinfo  {journal} {Nat. Phys.}\ }\textbf {\bibinfo {volume}
  {6}},\ \bibinfo {pages} {13} (\bibinfo {year} {2010})}\BibitemShut {NoStop}%
\bibitem [{\citenamefont {Amini}\ \emph {et~al.}(2010)\citenamefont {Amini},
  \citenamefont {Uys}, \citenamefont {Wesenberg}, \citenamefont {Seidelin},
  \citenamefont {Britton}, \citenamefont {Bollinger}, \citenamefont
  {Leibfried}, \citenamefont {Ospelkaus}, \citenamefont {{VanDevender}},\ and\
  \citenamefont {Wineland}}]{amini_toward_2010}%
  \BibitemOpen
  \bibfield  {author} {\bibinfo {author} {\bibfnamefont {J.~M.}\ \bibnamefont
  {Amini}}, \bibinfo {author} {\bibfnamefont {H.}~\bibnamefont {Uys}}, \bibinfo
  {author} {\bibfnamefont {J.~H.}\ \bibnamefont {Wesenberg}}, \bibinfo {author}
  {\bibfnamefont {S.}~\bibnamefont {Seidelin}}, \bibinfo {author}
  {\bibfnamefont {J.}~\bibnamefont {Britton}}, \bibinfo {author} {\bibfnamefont
  {J.~J.}\ \bibnamefont {Bollinger}}, \bibinfo {author} {\bibfnamefont
  {D.}~\bibnamefont {Leibfried}}, \bibinfo {author} {\bibfnamefont
  {C.}~\bibnamefont {Ospelkaus}}, \bibinfo {author} {\bibfnamefont {A.~P.}\
  \bibnamefont {{VanDevender}}}, \ and\ \bibinfo {author} {\bibfnamefont
  {D.~J.}\ \bibnamefont {Wineland}},\ }\href@noop {} {\bibfield  {journal}
  {\bibinfo  {journal} {New J. Phys.}\ }\textbf {\bibinfo {volume} {12}},\
  \bibinfo {pages} {033031} (\bibinfo {year} {2010})}\BibitemShut {NoStop}%
\bibitem [{\citenamefont {Moehring}\ \emph {et~al.}(2011)\citenamefont
  {Moehring}, \citenamefont {Highstrete}, \citenamefont {Stick}, \citenamefont
  {Fortier}, \citenamefont {Haltli}, \citenamefont {Tigges},\ and\
  \citenamefont {Blain}}]{moehring_design_2011}%
  \BibitemOpen
  \bibfield  {author} {\bibinfo {author} {\bibfnamefont {D.~L.}\ \bibnamefont
  {Moehring}}, \bibinfo {author} {\bibfnamefont {C.}~\bibnamefont
  {Highstrete}}, \bibinfo {author} {\bibfnamefont {D.}~\bibnamefont {Stick}},
  \bibinfo {author} {\bibfnamefont {K.~M.}\ \bibnamefont {Fortier}}, \bibinfo
  {author} {\bibfnamefont {R.}~\bibnamefont {Haltli}}, \bibinfo {author}
  {\bibfnamefont {C.}~\bibnamefont {Tigges}}, \ and\ \bibinfo {author}
  {\bibfnamefont {M.~G.}\ \bibnamefont {Blain}},\ }\href@noop {} {\bibfield
  {journal} {\bibinfo  {journal} {New J. Phys.}\ }\textbf {\bibinfo {volume}
  {13}},\ \bibinfo {pages} {075018} (\bibinfo {year} {2011})}\BibitemShut
  {NoStop}%
\bibitem [{\citenamefont {Doret}\ \emph {et~al.}(2012)\citenamefont {Doret},
  \citenamefont {Amini}, \citenamefont {Wright}, \citenamefont {Volin},
  \citenamefont {Killian}, \citenamefont {Ozakin}, \citenamefont {Denison},
  \citenamefont {Hayden}, \citenamefont {Pai}, \citenamefont {Slusher},\ and\
  \citenamefont {Harter}}]{doret_controlling_2012}%
  \BibitemOpen
  \bibfield  {author} {\bibinfo {author} {\bibfnamefont {S.}~\bibnamefont
  {Doret}}, \bibinfo {author} {\bibfnamefont {J.~M.}\ \bibnamefont {Amini}},
  \bibinfo {author} {\bibfnamefont {K.}~\bibnamefont {Wright}}, \bibinfo
  {author} {\bibfnamefont {C.}~\bibnamefont {Volin}}, \bibinfo {author}
  {\bibfnamefont {T.}~\bibnamefont {Killian}}, \bibinfo {author} {\bibfnamefont
  {A.}~\bibnamefont {Ozakin}}, \bibinfo {author} {\bibfnamefont
  {D.}~\bibnamefont {Denison}}, \bibinfo {author} {\bibfnamefont
  {H.}~\bibnamefont {Hayden}}, \bibinfo {author} {\bibfnamefont {C.-S.}\
  \bibnamefont {Pai}}, \bibinfo {author} {\bibfnamefont {R.~E.}\ \bibnamefont
  {Slusher}}, \ and\ \bibinfo {author} {\bibfnamefont {A.~W.}\ \bibnamefont
  {Harter}},\ }\href@noop {} {\bibfield  {journal} {\bibinfo  {journal} {New J.
  Phys.}\ }\textbf {\bibinfo {volume} {14}},\ \bibinfo {pages} {073012}
  (\bibinfo {year} {2012})}\BibitemShut {NoStop}%
\bibitem [{\citenamefont {Blakestad}\ \emph {et~al.}(2011)\citenamefont
  {Blakestad}, \citenamefont {Ospelkaus}, \citenamefont {{VanDevender}},
  \citenamefont {Wesenberg}, \citenamefont {Biercuk}, \citenamefont
  {Leibfried},\ and\ \citenamefont
  {Wineland}}]{blakestad_near-ground-state_2011}%
  \BibitemOpen
  \bibfield  {author} {\bibinfo {author} {\bibfnamefont {R.~B.}\ \bibnamefont
  {Blakestad}}, \bibinfo {author} {\bibfnamefont {C.}~\bibnamefont
  {Ospelkaus}}, \bibinfo {author} {\bibfnamefont {A.~P.}\ \bibnamefont
  {{VanDevender}}}, \bibinfo {author} {\bibfnamefont {J.~H.}\ \bibnamefont
  {Wesenberg}}, \bibinfo {author} {\bibfnamefont {M.~J.}\ \bibnamefont
  {Biercuk}}, \bibinfo {author} {\bibfnamefont {D.}~\bibnamefont {Leibfried}},
  \ and\ \bibinfo {author} {\bibfnamefont {D.~J.}\ \bibnamefont {Wineland}},\
  }\href@noop {} {\bibfield  {journal} {\bibinfo  {journal} {Phys. Rev. A}\
  }\textbf {\bibinfo {volume} {84}},\ \bibinfo {pages} {032314} (\bibinfo
  {year} {2011})}\BibitemShut {NoStop}%
\bibitem [{\citenamefont {Walther}\ \emph {et~al.}(2012)\citenamefont
  {Walther}, \citenamefont {Ziesel}, \citenamefont {Ruster}, \citenamefont
  {Dawkins}, \citenamefont {Ott}, \citenamefont {Hettrich}, \citenamefont
  {Singer}, \citenamefont {{Schmidt-Kaler}},\ and\ \citenamefont
  {Poschinger}}]{walther_controlling_2012}%
  \BibitemOpen
  \bibfield  {author} {\bibinfo {author} {\bibfnamefont {A.}~\bibnamefont
  {Walther}}, \bibinfo {author} {\bibfnamefont {F.}~\bibnamefont {Ziesel}},
  \bibinfo {author} {\bibfnamefont {T.}~\bibnamefont {Ruster}}, \bibinfo
  {author} {\bibfnamefont {S.~T.}\ \bibnamefont {Dawkins}}, \bibinfo {author}
  {\bibfnamefont {K.}~\bibnamefont {Ott}}, \bibinfo {author} {\bibfnamefont
  {M.}~\bibnamefont {Hettrich}}, \bibinfo {author} {\bibfnamefont
  {K.}~\bibnamefont {Singer}}, \bibinfo {author} {\bibfnamefont
  {F.}~\bibnamefont {{Schmidt-Kaler}}}, \ and\ \bibinfo {author} {\bibfnamefont
  {U.}~\bibnamefont {Poschinger}},\ }\href@noop {} {\bibfield  {journal}
  {\bibinfo  {journal} {Phys. Rev. Lett.}\ }\textbf {\bibinfo {volume} {109}},\
  \bibinfo {pages} {080501} (\bibinfo {year} {2012})}\BibitemShut {NoStop}%
\bibitem [{\citenamefont {Bowler}\ \emph {et~al.}(2012)\citenamefont {Bowler},
  \citenamefont {Gaebler}, \citenamefont {Lin}, \citenamefont {Tan},
  \citenamefont {Hanneke}, \citenamefont {Jost}, \citenamefont {Home},
  \citenamefont {Leibfried},\ and\ \citenamefont
  {Wineland}}]{bowler_coherent_2012}%
  \BibitemOpen
  \bibfield  {author} {\bibinfo {author} {\bibfnamefont {R.}~\bibnamefont
  {Bowler}}, \bibinfo {author} {\bibfnamefont {J.}~\bibnamefont {Gaebler}},
  \bibinfo {author} {\bibfnamefont {Y.}~\bibnamefont {Lin}}, \bibinfo {author}
  {\bibfnamefont {T.~R.}\ \bibnamefont {Tan}}, \bibinfo {author} {\bibfnamefont
  {D.}~\bibnamefont {Hanneke}}, \bibinfo {author} {\bibfnamefont {J.~D.}\
  \bibnamefont {Jost}}, \bibinfo {author} {\bibfnamefont {J.~P.}\ \bibnamefont
  {Home}}, \bibinfo {author} {\bibfnamefont {D.}~\bibnamefont {Leibfried}}, \
  and\ \bibinfo {author} {\bibfnamefont {D.~J.}\ \bibnamefont {Wineland}},\
  }\href@noop {} {\bibfield  {journal} {\bibinfo  {journal} {Phys. Rev. Lett.}\
  }\textbf {\bibinfo {volume} {109}},\ \bibinfo {pages} {080502} (\bibinfo
  {year} {2012})}\BibitemShut {NoStop}%
\bibitem [{\citenamefont {Wineland}\ \emph {et~al.}(1998)\citenamefont
  {Wineland}, \citenamefont {Monroe}, \citenamefont {Itano}, \citenamefont
  {Leibfried}, \citenamefont {King},\ and\ \citenamefont
  {Meekhof}}]{wineland_experimental_1998}%
  \BibitemOpen
  \bibfield  {author} {\bibinfo {author} {\bibfnamefont {D.~J.}\ \bibnamefont
  {Wineland}}, \bibinfo {author} {\bibfnamefont {C.}~\bibnamefont {Monroe}},
  \bibinfo {author} {\bibfnamefont {W.~M.}\ \bibnamefont {Itano}}, \bibinfo
  {author} {\bibfnamefont {D.}~\bibnamefont {Leibfried}}, \bibinfo {author}
  {\bibfnamefont {B.~E.}\ \bibnamefont {King}}, \ and\ \bibinfo {author}
  {\bibfnamefont {D.~M.}\ \bibnamefont {Meekhof}},\ }\href@noop {} {\bibfield
  {journal} {\bibinfo  {journal} {J. Res. Natl. Inst. Stand. Technol.}\
  }\textbf {\bibinfo {volume} {103}},\ \bibinfo {pages} {259} (\bibinfo {year}
  {1998})}\BibitemShut {NoStop}%
\bibitem [{\citenamefont {Kielpinski}\ \emph {et~al.}(2002)\citenamefont
  {Kielpinski}, \citenamefont {Monroe},\ and\ \citenamefont
  {Wineland}}]{kielpinski_architecture_2002}%
  \BibitemOpen
  \bibfield  {author} {\bibinfo {author} {\bibfnamefont {D.}~\bibnamefont
  {Kielpinski}}, \bibinfo {author} {\bibfnamefont {C.}~\bibnamefont {Monroe}},
  \ and\ \bibinfo {author} {\bibfnamefont {D.~J.}\ \bibnamefont {Wineland}},\
  }\href@noop {} {\bibfield  {journal} {\bibinfo  {journal} {Nature}\ }\textbf
  {\bibinfo {volume} {417}},\ \bibinfo {pages} {709} (\bibinfo {year}
  {2002})}\BibitemShut {NoStop}%
\bibitem [{\citenamefont {Mintert}\ and\ \citenamefont
  {Wunderlich}(2001)}]{mintert_ion-trap_2001}%
  \BibitemOpen
  \bibfield  {author} {\bibinfo {author} {\bibfnamefont {F.}~\bibnamefont
  {Mintert}}\ and\ \bibinfo {author} {\bibfnamefont {C.}~\bibnamefont
  {Wunderlich}},\ }\href@noop {} {\bibfield  {journal} {\bibinfo  {journal}
  {Phys. Rev. Lett.}\ }\textbf {\bibinfo {volume} {87}},\ \bibinfo {pages}
  {257904} (\bibinfo {year} {2001})}\BibitemShut {NoStop}%
\bibitem [{\citenamefont {Chiaverini}\ and\ \citenamefont
  {Lybarger}(2008)}]{chiaverini_laserless_2008}%
  \BibitemOpen
  \bibfield  {author} {\bibinfo {author} {\bibfnamefont {J.}~\bibnamefont
  {Chiaverini}}\ and\ \bibinfo {author} {\bibfnamefont {W.~E.}\ \bibnamefont
  {Lybarger}},\ }\href@noop {} {\bibfield  {journal} {\bibinfo  {journal}
  {Phys. Rev. A}\ }\textbf {\bibinfo {volume} {77}},\ \bibinfo {pages} {022324}
  (\bibinfo {year} {2008})}\BibitemShut {NoStop}%
\bibitem [{\citenamefont {Ospelkaus}\ \emph {et~al.}(2008)\citenamefont
  {Ospelkaus}, \citenamefont {Langer}, \citenamefont {Amini}, \citenamefont
  {Brown}, \citenamefont {Leibfried},\ and\ \citenamefont
  {Wineland}}]{ospelkaus_trapped-ion_2008}%
  \BibitemOpen
  \bibfield  {author} {\bibinfo {author} {\bibfnamefont {C.}~\bibnamefont
  {Ospelkaus}}, \bibinfo {author} {\bibfnamefont {C.~E.}\ \bibnamefont
  {Langer}}, \bibinfo {author} {\bibfnamefont {J.~M.}\ \bibnamefont {Amini}},
  \bibinfo {author} {\bibfnamefont {K.~R.}\ \bibnamefont {Brown}}, \bibinfo
  {author} {\bibfnamefont {D.}~\bibnamefont {Leibfried}}, \ and\ \bibinfo
  {author} {\bibfnamefont {D.~J.}\ \bibnamefont {Wineland}},\ }\href@noop {}
  {\bibfield  {journal} {\bibinfo  {journal} {Phys. Rev. Lett.}\ }\textbf
  {\bibinfo {volume} {101}},\ \bibinfo {pages} {090502} (\bibinfo {year}
  {2008})}\BibitemShut {NoStop}%
\bibitem [{\citenamefont {Johanning}\ \emph {et~al.}(2009)\citenamefont
  {Johanning}, \citenamefont {Braun}, \citenamefont {Timoney}, \citenamefont
  {Elman}, \citenamefont {Neuhauser},\ and\ \citenamefont
  {Wunderlich}}]{johanning_individual_2009}%
  \BibitemOpen
  \bibfield  {author} {\bibinfo {author} {\bibfnamefont {M.}~\bibnamefont
  {Johanning}}, \bibinfo {author} {\bibfnamefont {A.}~\bibnamefont {Braun}},
  \bibinfo {author} {\bibfnamefont {N.}~\bibnamefont {Timoney}}, \bibinfo
  {author} {\bibfnamefont {V.}~\bibnamefont {Elman}}, \bibinfo {author}
  {\bibfnamefont {W.}~\bibnamefont {Neuhauser}}, \ and\ \bibinfo {author}
  {\bibfnamefont {C.}~\bibnamefont {Wunderlich}},\ }\href@noop {} {\bibfield
  {journal} {\bibinfo  {journal} {Phys. Rev. Lett.}\ }\textbf {\bibinfo
  {volume} {102}},\ \bibinfo {pages} {073004} (\bibinfo {year}
  {2009})}\BibitemShut {NoStop}%
\bibitem [{\citenamefont {Wang}\ \emph {et~al.}(2009)\citenamefont {Wang},
  \citenamefont {Labaziewicz}, \citenamefont {Ge}, \citenamefont {Shewmon},\
  and\ \citenamefont {Chuang}}]{wang_individual_2009}%
  \BibitemOpen
  \bibfield  {author} {\bibinfo {author} {\bibfnamefont {S.~X.}\ \bibnamefont
  {Wang}}, \bibinfo {author} {\bibfnamefont {J.}~\bibnamefont {Labaziewicz}},
  \bibinfo {author} {\bibfnamefont {Y.}~\bibnamefont {Ge}}, \bibinfo {author}
  {\bibfnamefont {R.}~\bibnamefont {Shewmon}}, \ and\ \bibinfo {author}
  {\bibfnamefont {I.~L.}\ \bibnamefont {Chuang}},\ }\href@noop {} {\bibfield
  {journal} {\bibinfo  {journal} {Appl. Phys. Lett.}\ }\textbf {\bibinfo
  {volume} {94}},\ \bibinfo {pages} {094103} (\bibinfo {year}
  {2009})}\BibitemShut {NoStop}%
\bibitem [{\citenamefont {Khromova}\ \emph {et~al.}(2012)\citenamefont
  {Khromova}, \citenamefont {Piltz}, \citenamefont {Scharfenberger},
  \citenamefont {Gloger}, \citenamefont {Johanning}, \citenamefont
  {Var\'{o}n},\ and\ \citenamefont {Wunderlich}}]{khromova_designer_2012}%
  \BibitemOpen
  \bibfield  {author} {\bibinfo {author} {\bibfnamefont {A.}~\bibnamefont
  {Khromova}}, \bibinfo {author} {\bibfnamefont {C.}~\bibnamefont {Piltz}},
  \bibinfo {author} {\bibfnamefont {B.}~\bibnamefont {Scharfenberger}},
  \bibinfo {author} {\bibfnamefont {T.~F.}\ \bibnamefont {Gloger}}, \bibinfo
  {author} {\bibfnamefont {M.}~\bibnamefont {Johanning}}, \bibinfo {author}
  {\bibfnamefont {A.~F.}\ \bibnamefont {Var\'{o}n}}, \ and\ \bibinfo {author}
  {\bibfnamefont {C.}~\bibnamefont {Wunderlich}},\ }\href@noop {} {\bibfield
  {journal} {\bibinfo  {journal} {Phys. Rev. Lett.}\ }\textbf {\bibinfo
  {volume} {108}},\ \bibinfo {pages} {220502} (\bibinfo {year}
  {2012})}\BibitemShut {NoStop}%
\bibitem [{\citenamefont {Brown}\ \emph {et~al.}(2011)\citenamefont {Brown},
  \citenamefont {Wilson}, \citenamefont {Colombe}, \citenamefont {Ospelkaus},
  \citenamefont {Meier}, \citenamefont {Knill}, \citenamefont {Leibfried},\
  and\ \citenamefont {Wineland}}]{brown_single-qubit-gate_2011}%
  \BibitemOpen
  \bibfield  {author} {\bibinfo {author} {\bibfnamefont {K.~R.}\ \bibnamefont
  {Brown}}, \bibinfo {author} {\bibfnamefont {A.~C.}\ \bibnamefont {Wilson}},
  \bibinfo {author} {\bibfnamefont {Y.}~\bibnamefont {Colombe}}, \bibinfo
  {author} {\bibfnamefont {C.}~\bibnamefont {Ospelkaus}}, \bibinfo {author}
  {\bibfnamefont {A.~M.}\ \bibnamefont {Meier}}, \bibinfo {author}
  {\bibfnamefont {E.}~\bibnamefont {Knill}}, \bibinfo {author} {\bibfnamefont
  {D.}~\bibnamefont {Leibfried}}, \ and\ \bibinfo {author} {\bibfnamefont
  {D.~J.}\ \bibnamefont {Wineland}},\ }\href@noop {} {\bibfield  {journal}
  {\bibinfo  {journal} {Phys. Rev. A}\ }\textbf {\bibinfo {volume} {84}},\
  \bibinfo {pages} {R030303} (\bibinfo {year} {2011})}\BibitemShut {NoStop}%
\bibitem [{\citenamefont {Ospelkaus}\ \emph {et~al.}(2011)\citenamefont
  {Ospelkaus}, \citenamefont {Warring}, \citenamefont {Colombe}, \citenamefont
  {Brown}, \citenamefont {Amini}, \citenamefont {Leibfried},\ and\
  \citenamefont {Wineland}}]{ospelkaus_microwave_2011}%
  \BibitemOpen
  \bibfield  {author} {\bibinfo {author} {\bibfnamefont {C.}~\bibnamefont
  {Ospelkaus}}, \bibinfo {author} {\bibfnamefont {U.}~\bibnamefont {Warring}},
  \bibinfo {author} {\bibfnamefont {Y.}~\bibnamefont {Colombe}}, \bibinfo
  {author} {\bibfnamefont {K.~R.}\ \bibnamefont {Brown}}, \bibinfo {author}
  {\bibfnamefont {J.~M.}\ \bibnamefont {Amini}}, \bibinfo {author}
  {\bibfnamefont {D.}~\bibnamefont {Leibfried}}, \ and\ \bibinfo {author}
  {\bibfnamefont {D.~J.}\ \bibnamefont {Wineland}},\ }\href@noop {} {\bibfield
  {journal} {\bibinfo  {journal} {Nature}\ }\textbf {\bibinfo {volume} {476}},\
  \bibinfo {pages} {181} (\bibinfo {year} {2011})}\BibitemShut {NoStop}%
\bibitem [{\citenamefont {Warring}\ \emph {et~al.}(2012)\citenamefont
  {Warring}, \citenamefont {Ospelkaus}, \citenamefont {Colombe}, \citenamefont
  {Brown}, \citenamefont {Amini}, \citenamefont {Carsjens}, \citenamefont
  {Leibfried},\ and\ \citenamefont {Wineland}}]{warring_microwave_2012}%
  \BibitemOpen
  \bibfield  {author} {\bibinfo {author} {\bibfnamefont {U.}~\bibnamefont
  {Warring}}, \bibinfo {author} {\bibfnamefont {C.}~\bibnamefont {Ospelkaus}},
  \bibinfo {author} {\bibfnamefont {Y.}~\bibnamefont {Colombe}}, \bibinfo
  {author} {\bibfnamefont {K.~R.}\ \bibnamefont {Brown}}, \bibinfo {author}
  {\bibfnamefont {J.~M.}\ \bibnamefont {Amini}}, \bibinfo {author}
  {\bibfnamefont {M.}~\bibnamefont {Carsjens}}, \bibinfo {author}
  {\bibfnamefont {D.}~\bibnamefont {Leibfried}}, \ and\ \bibinfo {author}
  {\bibfnamefont {D.~J.}\ \bibnamefont {Wineland}},\ }\href@noop {} {\bibfield
  {journal} {\bibinfo  {journal} {in preparation}\ } (\bibinfo {year}
  {2012})}\BibitemShut {NoStop}%
\bibitem [{\citenamefont {Allcock}\ \emph {et~al.}(2012)\citenamefont
  {Allcock}, \citenamefont {Harty}, \citenamefont {Ballance}, \citenamefont
  {Keitch}, \citenamefont {Linke}, \citenamefont {Stacey},\ and\ \citenamefont
  {Lucas}}]{allcock_microfabricated_2012}%
  \BibitemOpen
  \bibfield  {author} {\bibinfo {author} {\bibfnamefont {D.}~\bibnamefont
  {Allcock}}, \bibinfo {author} {\bibfnamefont {T.}~\bibnamefont {Harty}},
  \bibinfo {author} {\bibfnamefont {C.~J.}\ \bibnamefont {Ballance}}, \bibinfo
  {author} {\bibfnamefont {B.~C.}\ \bibnamefont {Keitch}}, \bibinfo {author}
  {\bibfnamefont {N.~M.}\ \bibnamefont {Linke}}, \bibinfo {author}
  {\bibfnamefont {D.~N.}\ \bibnamefont {Stacey}}, \ and\ \bibinfo {author}
  {\bibfnamefont {D.~M.}\ \bibnamefont {Lucas}},\ }\href@noop {} {\bibfield
  {journal} {\bibinfo  {journal} {arXiv:1210.3272}\ } (\bibinfo {year}
  {2012})}\BibitemShut {NoStop}%
\bibitem [{\citenamefont {N\"agerl}\ \emph {et~al.}(1999)\citenamefont
  {N\"agerl}, \citenamefont {Leibfried}, \citenamefont {Rohde}, \citenamefont
  {Thalhammer}, \citenamefont {Eschner}, \citenamefont {{Schmidt-Kaler}},\ and\
  \citenamefont {Blatt}}]{naegerl_laser_1999}%
  \BibitemOpen
  \bibfield  {author} {\bibinfo {author} {\bibfnamefont {H.~C.}\ \bibnamefont
  {N\"agerl}}, \bibinfo {author} {\bibfnamefont {D.}~\bibnamefont {Leibfried}},
  \bibinfo {author} {\bibfnamefont {H.}~\bibnamefont {Rohde}}, \bibinfo
  {author} {\bibfnamefont {G.}~\bibnamefont {Thalhammer}}, \bibinfo {author}
  {\bibfnamefont {J.}~\bibnamefont {Eschner}}, \bibinfo {author} {\bibfnamefont
  {F.}~\bibnamefont {{Schmidt-Kaler}}}, \ and\ \bibinfo {author} {\bibfnamefont
  {R.}~\bibnamefont {Blatt}},\ }\href@noop {} {\bibfield  {journal} {\bibinfo
  {journal} {Phys. Rev. A}\ }\textbf {\bibinfo {volume} {60}},\ \bibinfo
  {pages} {145} (\bibinfo {year} {1999})}\BibitemShut {NoStop}%
\bibitem [{\citenamefont {Rowe}\ \emph {et~al.}(2001)\citenamefont {Rowe},
  \citenamefont {Kielpinski}, \citenamefont {Meyer}, \citenamefont {Sackett},
  \citenamefont {Itano}, \citenamefont {Monroe},\ and\ \citenamefont
  {Wineland}}]{rowe_experimental_2001}%
  \BibitemOpen
  \bibfield  {author} {\bibinfo {author} {\bibfnamefont {M.~A.}\ \bibnamefont
  {Rowe}}, \bibinfo {author} {\bibfnamefont {D.}~\bibnamefont {Kielpinski}},
  \bibinfo {author} {\bibfnamefont {V.}~\bibnamefont {Meyer}}, \bibinfo
  {author} {\bibfnamefont {C.~A.}\ \bibnamefont {Sackett}}, \bibinfo {author}
  {\bibfnamefont {W.~M.}\ \bibnamefont {Itano}}, \bibinfo {author}
  {\bibfnamefont {C.}~\bibnamefont {Monroe}}, \ and\ \bibinfo {author}
  {\bibfnamefont {D.~J.}\ \bibnamefont {Wineland}},\ }\href@noop {} {\bibfield
  {journal} {\bibinfo  {journal} {Nature}\ }\textbf {\bibinfo {volume} {409}},\
  \bibinfo {pages} {791} (\bibinfo {year} {2001})}\BibitemShut {NoStop}%
\bibitem [{\citenamefont {Schrader}\ \emph {et~al.}(2004)\citenamefont
  {Schrader}, \citenamefont {Dotsenko}, \citenamefont {Khudaverdyan},
  \citenamefont {Miroshnychenko}, \citenamefont {Rauschenbeutel},\ and\
  \citenamefont {Meschede}}]{schrader_neutral_2004}%
  \BibitemOpen
  \bibfield  {author} {\bibinfo {author} {\bibfnamefont {D.}~\bibnamefont
  {Schrader}}, \bibinfo {author} {\bibfnamefont {I.}~\bibnamefont {Dotsenko}},
  \bibinfo {author} {\bibfnamefont {M.}~\bibnamefont {Khudaverdyan}}, \bibinfo
  {author} {\bibfnamefont {Y.}~\bibnamefont {Miroshnychenko}}, \bibinfo
  {author} {\bibfnamefont {A.}~\bibnamefont {Rauschenbeutel}}, \ and\ \bibinfo
  {author} {\bibfnamefont {D.}~\bibnamefont {Meschede}},\ }\href@noop {}
  {\bibfield  {journal} {\bibinfo  {journal} {Phys. Rev. Lett.}\ }\textbf
  {\bibinfo {volume} {93}},\ \bibinfo {pages} {150501} (\bibinfo {year}
  {2004})}\BibitemShut {NoStop}%
\bibitem [{\citenamefont {Seidelin}\ \emph {et~al.}(2006)\citenamefont
  {Seidelin} \emph {et~al.}}]{seidelin_microfabricated_2006}%
  \BibitemOpen
  \bibfield  {author} {\bibinfo {author} {\bibfnamefont {S.}~\bibnamefont
  {Seidelin}} \emph {et~al.},\ }\href@noop {} {\bibfield  {journal} {\bibinfo
  {journal} {Phys. Rev. Lett.}\ }\textbf {\bibinfo {volume} {96}},\ \bibinfo
  {pages} {253003} (\bibinfo {year} {2006})}\BibitemShut {NoStop}%
\bibitem [{Note1()}]{Note1}%
  \BibitemOpen
  \bibinfo {note} {Here $F$ is the total angular momentum and $m_{\protect
  \text {F}}$ is the projection of the angular momentum onto the magnetic
  field axis}\BibitemShut {NoStop}%
\bibitem [{\citenamefont {Langer}\ \emph {et~al.}(2005)\citenamefont {Langer}
  \emph {et~al.}}]{langer_long-lived_2005}%
  \BibitemOpen
  \bibfield  {author} {\bibinfo {author} {\bibfnamefont {C.}~\bibnamefont
  {Langer}} \emph {et~al.},\ }\href@noop {} {\bibfield  {journal} {\bibinfo
  {journal} {Phys. Rev. Lett.}\ }\textbf {\bibinfo {volume} {95}},\ \bibinfo
  {pages} {060502} (\bibinfo {year} {2005})}\BibitemShut {NoStop}%
\bibitem [{\citenamefont {Turchette}\ \emph {et~al.}(1998)\citenamefont
  {Turchette}, \citenamefont {Wood}, \citenamefont {King}, \citenamefont
  {Myatt}, \citenamefont {Leibfried}, \citenamefont {Itano}, \citenamefont
  {Monroe},\ and\ \citenamefont {Wineland}}]{turchette_deterministic_1998}%
  \BibitemOpen
  \bibfield  {author} {\bibinfo {author} {\bibfnamefont {Q.~A.}\ \bibnamefont
  {Turchette}}, \bibinfo {author} {\bibfnamefont {C.~S.}\ \bibnamefont {Wood}},
  \bibinfo {author} {\bibfnamefont {B.~E.}\ \bibnamefont {King}}, \bibinfo
  {author} {\bibfnamefont {C.~J.}\ \bibnamefont {Myatt}}, \bibinfo {author}
  {\bibfnamefont {D.}~\bibnamefont {Leibfried}}, \bibinfo {author}
  {\bibfnamefont {W.~M.}\ \bibnamefont {Itano}}, \bibinfo {author}
  {\bibfnamefont {C.}~\bibnamefont {Monroe}}, \ and\ \bibinfo {author}
  {\bibfnamefont {D.~J.}\ \bibnamefont {Wineland}},\ }\href@noop {} {\bibfield
  {journal} {\bibinfo  {journal} {Phys. Rev. Lett.}\ }\textbf {\bibinfo
  {volume} {81}},\ \bibinfo {pages} {3631{\textendash}3634} (\bibinfo {year}
  {1998})}\BibitemShut {NoStop}%
\bibitem [{\citenamefont {Leibfried}(1999)}]{leibfried_individual_1999}%
  \BibitemOpen
  \bibfield  {author} {\bibinfo {author} {\bibfnamefont {D.}~\bibnamefont
  {Leibfried}},\ }\href@noop {} {\bibfield  {journal} {\bibinfo  {journal}
  {Phys. Rev. A}\ }\textbf {\bibinfo {volume} {60}},\ \bibinfo {pages} {R3335}
  (\bibinfo {year} {1999})}\BibitemShut {NoStop}%
\bibitem [{\citenamefont {Berkeland}\ \emph {et~al.}(1998)\citenamefont
  {Berkeland}, \citenamefont {Miller}, \citenamefont {Bergquist}, \citenamefont
  {Itano},\ and\ \citenamefont {Wineland}}]{berkeland_minimization_1998}%
  \BibitemOpen
  \bibfield  {author} {\bibinfo {author} {\bibfnamefont {D.~J.}\ \bibnamefont
  {Berkeland}}, \bibinfo {author} {\bibfnamefont {J.~D.}\ \bibnamefont
  {Miller}}, \bibinfo {author} {\bibfnamefont {J.~C.}\ \bibnamefont
  {Bergquist}}, \bibinfo {author} {\bibfnamefont {W.~M.}\ \bibnamefont
  {Itano}}, \ and\ \bibinfo {author} {\bibfnamefont {D.~J.}\ \bibnamefont
  {Wineland}},\ }\href@noop {} {\bibfield  {journal} {\bibinfo  {journal} {J.
  Appl. Phys.}\ }\textbf {\bibinfo {volume} {83}},\ \bibinfo {pages} {5025}
  (\bibinfo {year} {1998})}\BibitemShut {NoStop}%
\bibitem [{Note2()}]{Note2}%
  \BibitemOpen
  \bibinfo {note} {The coherence time is defined as the Ramsey interval
  $T_{\protect \text {R}}$ at which the fringe contrast decays by a factor
  $e^{-1}$.}\BibitemShut {Stop}%
\bibitem [{\citenamefont {Levitt}(1986)}]{levitt_composite_1986}%
  \BibitemOpen
  \bibfield  {author} {\bibinfo {author} {\bibfnamefont {M.}~\bibnamefont
  {Levitt}},\ }\href@noop {} {\bibfield  {journal} {\bibinfo  {journal} {Prog.
  Nucl. Magn. Reson. Spectrosc.}\ }\textbf {\bibinfo {volume} {18}},\ \bibinfo
  {pages} {61} (\bibinfo {year} {1986})}\BibitemShut {NoStop}%
\bibitem [{\citenamefont {Navon}\ \emph {et~al.}(2012)\citenamefont {Navon},
  \citenamefont {Kotler}, \citenamefont {Akerman}, \citenamefont {Glickman},
  \citenamefont {Almog},\ and\ \citenamefont {Ozeri}}]{navon_single-spin_2012}%
  \BibitemOpen
  \bibfield  {author} {\bibinfo {author} {\bibfnamefont {N.}~\bibnamefont
  {Navon}}, \bibinfo {author} {\bibfnamefont {S.}~\bibnamefont {Kotler}},
  \bibinfo {author} {\bibfnamefont {N.}~\bibnamefont {Akerman}}, \bibinfo
  {author} {\bibfnamefont {Y.}~\bibnamefont {Glickman}}, \bibinfo {author}
  {\bibfnamefont {I.}~\bibnamefont {Almog}}, \ and\ \bibinfo {author}
  {\bibfnamefont {R.}~\bibnamefont {Ozeri}},\ }\href@noop {} {\bibfield
  {journal} {\bibinfo  {journal} {arXiv:1210.7336}\ } (\bibinfo {year}
  {2012})}\BibitemShut {NoStop}%
\end{thebibliography}
\end{document}